\documentclass[twocolumn,showpacs,preprintnumbers,amsmath,amssymb]{revtex4}
\usepackage{graphicx,epsfig}
\usepackage{dcolumn}
\usepackage{bm}
\newcommand{\be}{\begin{equation}}
\newcommand{\ee}{\end{equation}}
\newcommand{\bse}{\begin{subequations}}
\newcommand{\ese}{\end{subequations}}
\newcommand{\bary}{\begin{eqnarray}}
\newcommand{\eary}{\end{eqnarray}}

\begin{document}

\title{Can there be neutrino oscillation in Gamma-Ray Bursts fireball ?}
\author{Sarira Sahu and Juan Carlos D'Olivo}

\affiliation{
Instituto de Ciencias Nucleares, Universidad Nacional Aut\'onoma de M\'exico, 
Circuito Exterior, C.U., A. Postal 70-543, 04510 Mexico DF, Mexico\\
}

\begin{abstract}
 
The central engine which powers the Gamma-Ray Burst (GRB) fireball,
produces neutrinos in the energy range of about 5-20 MeV. 
Fractions of these neutrinos may propagate 
through the fireball which is far away from the central engine. 
We have studied the propagation of these neutrinos through the fireball
which is contaminated by baryons and have shown that, resonant
conversion of neutrinos are possible for the oscillations of 
$\nu_e\leftrightarrow \nu_{\mu,\tau}$,
$\nu_e\leftrightarrow \nu_{s}$ and 
${\bar\nu}_{\mu,\tau}\leftrightarrow {\bar\nu}_{s}$ if
the neutrino mass square difference and mixing angle are in the atmospheric
and/or LSND range. On the other hand  it is probably difficult for neutrinos
to have resonant oscillation if the neutrino parameters are in the solar 
neutrino range. From the resonance condition we have estimated the fireball 
temperature and the baryon load in it.

\end{abstract}

\pacs{14.60.Pq,98.70.Rz}
\maketitle

GRBs are short, non-thermal bursts of low energy 
($\sim$ 100 KeV-1 MeV) photons and release about $10^{51}$-$10^{53}$ ergs in a 
few seconds making them the most luminous object in the universe. A class 
of models call {\it fireball model} seems to explain the temporal structure
of the bursts and the non-thermal nature of their 
spectra\cite{piran1,waxman1,piran2,mes1,mes2}. 
Sudden release of copious amount of 
$\gamma$ rays into a compact region with a
size $c\delta t\sim 100$-$1000$ Km\cite{piran1} 
creates an opaque $\gamma-e^{\pm}$ fireball
due to the process $\gamma+\gamma\rightarrow e^+ + e^-$. The average optical
depth of this process\cite{guil} is 
$\tau_{\gamma\gamma}\simeq 10^{13}$.
This optical depth is very large and even if there are no pairs to 
begin with, they will form very 
rapidly and will Compton scatter lower energy photons. Because of the huge
optical depth, photons can not escape freely. In the fireball the $\gamma$ and
$e^{\pm}$ pairs will thermalize with a temperature of about 3-10 MeV.
The fireball  expands relativistically with a Lorentz factor 
$\Gamma\sim 100-1000$  under its
own pressure and cools adiabatically due to the expansion. The radiation 
emerges freely to the inter galactic medium (ISM), 
when the optical depth is $\tau_{\gamma\gamma}\simeq 1$. 

In addition to $\gamma$, $e^{\pm}$ pairs, fireballs may also contain some
baryons, both from the progenitor and the surrounding medium.
These baryons can be either free or in the form of nuclei. 
If the fireball temperature
is high enough (more than 0.7 MeV), then it will be mostly in the form of 
neutrons and protons. Derishev et al. \cite{deri1,deri2}
argued that roughly equal numbers of 
neutrons and protons should be present in the fireball.
The electrons associated with the matter (baryons) can 
increase the opacity, hence delaying the process of emission of radiation and
the baryons can be accelerated along with the fireball and convert part of the
radiation energy into bulk kinetic energy.
But irrespective of it, the baryonic load has to be very 
small, otherwise, the expansion of the fireball will be
Newtonian, which is inconsistent with the present observations. Why the 
baryonic loading is so low in the fireball is still an open question to be 
answered\cite{piran1}. The neutrino oscillation may overcome the baryon loading
problem\cite{klu,vol} ! The evolution of the pure fireball (with no baryons) 
has been studied in ref.\cite{pac,good}. In the expanding fireball, protons 
are accelerated in shocks and collide with the photons to produce charged
pions, which give rise to ultra high-energy neutrinos\cite{waxman}. 
These neutrinos can be detected by $km^2$ detectors. Observation of these 
neutrinos can study oscillating neutrino flavors in the largest possible 
baseline, test the equivalence principle and many other neutrino properties
\cite{weiler}.

The hidden central engine which powers the fireball is still unknown, but 
observation suggests that it must be compact. The prime candidates 
are merger of neutron star with neutron star (NS-NS),
black hole-neutron star binaries (BH-NS), hypernova/collapsar models 
involving a massive stellar progenitor\cite{piran1,waxman1,piran2,mes1,ruf}. 
The recent observations of GRB030329 strongly favor collapsar model
\cite{nature}. In all these models, the gravitational
energy is released mostly in the form of $\nu{\bar\nu}$, gravitational
radiation and a small fraction ($\sim 10^{-3}$) is responsible to power
GRB. Neutrinos
of energy about 5-20 MeV are generated due to the stellar collapse or
merger event that trigger the burst. Also due to nucleonic bremsstrahlung 
$NN\rightarrow NN\nu {\bar\nu}$ and $e^+e^-\rightarrow\nu {\bar\nu}$ processes
muon and tau types neutrinos can be produced\cite{raf2} 
during the merger process and its flux will be very small.
Fractions of these neutrinos may propagate through the fireball which is far 
away from the central engine.
Also  within the fireball, because of the weak interaction process 
$p+e^-\rightarrow n+ \nu_e$, MeV neutrinos can be generated and propagate
through it. The fireball plasma being in an extreme condition, may affect
the propagation of MeV neutrinos through it. 

In a heat bath the dispersion relation which governs the propagation of the
particle gets modified and this can have a drastic effect on the particle
propagation in the heat bath.
The neutrino propagation in a thermal bath has been studied 
extensively\cite{raf1,enq1,jc1,sahu1}. 
In general, the propagating neutrino will experience an
effective potential due to the particles in the heat bath.
Particularly the neutrino
propagation in the early universe hot plasma, as well as in the supernova 
medium has profound implications in the respective physics. For example
the neutrino oscillation in the early universe hot plasma may change the 
relative abundances of $\nu_e$ and ${\bar\nu}_e$ and affect the primordial
nucleosynthesis of light elements. In the supernova case, 
the neutrino oscillation in the dense and compact
environment can affect the cooling mechanism. 

The electron type
neutrinos have charge current as well as neutral current interactions, 
but the muon and tau types will experience only the neutral current 
interactions. To the leading order the effective potential experience by 
the neutrinos is proportional
to the difference of the particle anti-particle number densities. So if the
system under consideration has equal number of particles and anti-particles,
the leading order contribution vanishes.
On the other hand, the next-to-leading order contribution (i.e. the term
proportional to $1/M^4$, where M is the vector boson mass) is
proportional to the sum of the particle and anti-particle number densities
and this is the leading contribution to the effective potential. 
The early universe hot plasma, supposed to have almost equal population of
particles and anti-particles, hence
the leading contribution to the neutrino effective potential will be 
proportional to $1/M^4$. Similar situation can arise for neutrinos 
propagating in the GRB fireball if one consider mostly photon-lepton 
fireball\cite{piran1,piran2,mes1}. 
Here in the present work we consider a fireball containing mostly 
photon-lepton with little baryon contamination, which mimics the 
early universe hot plasma and study the neutrino 
propagation within it.

In a  relativistic and nondegenerate $e^{\pm}$, proton and neutron 
plasma, the effective
potential experiences by $\nu_e$ is \cite{enq1,jc1}   
\bary
V_{\nu_e}\simeq \sqrt{2} G_F N_{\gamma}
\left [{\cal L}_e 
-\left ( \frac{7\xi(4)}{\xi(3)}\right )^2 \frac{T^2}{M^2_W}
\right ],
\label{potnue}
\eary
and by a muon or tau neutrino it is given by
\be
V_{\nu_{\mu,\tau}}\simeq\sqrt{2} G_F N_{\gamma}
{\cal L}_{\mu,\tau},
\label{potnumt}
\ee
where 
\be
{\cal L}_e=\left(\frac{1}{2} + 2\sin^2\theta_W\right ) L_e 
+\left(\frac{1}{2} - 2\sin^2\theta_W\right) L_p-\frac{L_n}{2}, 
\ee
and
\be
{\cal L}_{\mu,\tau}=\left( -\frac{1}{2} + 2\sin^2\theta_W\right) (L_e-L_p)
-\frac{L_n}{2}.
\ee
The asymmetry of the particle $a$ is defined as
\be
L_a = \frac{(N_a-{\bar N}_a)}{N_{\gamma}},
\label{asym}
\ee
where $N_{\gamma} = \frac{2}{\pi^2}\xi(3) T^3$ is the number density of 
photons.  
For the anti-neutrinos, the effective potential will be given by changing
$L_a\rightarrow -L_a$. For a pure $\gamma$ and 
$e^{\pm}$ fireball, we have $L_p=L_n=0$.

We consider the neutrino oscillation processes:
$\nu_{e}\leftrightarrow \nu_{\mu,\tau}$,
$\nu_{e}\leftrightarrow \nu_s$ and the antineutrino processes. The effective
potential difference for 
$\nu_{e}\leftrightarrow \nu_{\mu,\tau}$ and
${\bar\nu}_{e}\leftrightarrow {\bar\nu}_{\mu,\tau}$ processes, is given by
\be
V \simeq 
4.02\times 10^{-12} T^3_{MeV} 
\left [ \pm L_e - 6.14\times 10^{-9} T^2_{MeV}\right ]~MeV,
\label{numericalpot}
\ee
where $\pm$ corresponds to $\nu$ and ${\bar\nu}$ respectively, and 
for $\nu_{e}\leftrightarrow \nu_s$ oscillation, the effective potential 
difference is given by Eq.(\ref{potnue}).
For pure $\gamma$, $e^{\pm}$ fireball (CP-symmetric), $L_e=0$ and only the
higher order term will contribute. 
The conversion probability for the above processes for a constant $V$ 
is given by,
\be
{\cal P}(t)=\frac{\Delta^2\sin^2 2\theta}{\omega^2} 
\sin^2({\frac{\omega t}{2}}),
\label{prob}
\ee
 where
\be
\omega={\sqrt{(V-\Delta\cos 2\theta)^2 + \Delta^2 \sin^2 2\theta}},
\label{wf}
\ee
with $\Delta=\Delta m^2/2E_{\nu}$, $V$ is the potential 
difference,
$E_{\nu}$ is the neutrino energy and $\theta$ is the neutrino
mixing angle. The oscillation length is given by
\be
L_{osc}=\frac{L_v}{\sqrt{\cos^2 2\theta (1-\frac{V}{\Delta \cos 2\theta})^2 +
\sin^2 2\theta}},
\ee
where $L_v={2\pi}/{\Delta}$ is the vacuum oscillation length which can be 
recovered for $V=0$.
For resonance to occur we should have from Eq.(\ref{wf})
\be
V=\Delta \cos 2\theta.
\label{reso}
\ee
We have so far assumed that lepton asymmetry in the fireball does not vary with
distance. But in reality, the lepton asymmetry changes with distance. So in
this case we will consider the adiabaticity of the resonant conversion. The
adiabaticity condition at the resonance\cite{smir} is given by
\be
\kappa_r\equiv 2.0\times 10^{-3} 
  \left ( \frac{{\tilde {\Delta m^2}} \sin 2\theta}{E_{MeV}}\right )^2
\frac{l_{cm}}{T^3_{MeV}} \left ( \frac{dL_e}{dx}\right )^{-1} \geq 1,
\label{adbi}
\ee
where $l_{cm}$ is some length scale expressed in centimeter, $x$ is a
dimensionless variable, ${\tilde {\Delta m^2}}$ is in unit of $eV^2$ 
and $T_{MeV}$ and $E_{MeV}$ are expressed in unit of MeV.
The above condition depends on the neutrino parameters and the change in 
the $L_e$ as one goes away  from the center of the fireball.

As already stated in the introduction,
baryon loading is still an outstanding problem. It is believed that
the contamination is very low 
$(10^{-8} M_{\odot}-10^{-5} M_{\odot})$\cite{piran1,waxman1}, so that, 
the fireball can have a 
ultra-relativistic expansion.  For simplicity, we consider a charge 
neutral spherical fireball ($L_e=L_p$) of initial radius $R$ with equal 
number of protons and neutrons in it.
Then the baryon load in the fireball is 
\be
M_{b}\sim 2.23\times 10^{-4} 
R^3_7 T^3_{MeV} L_e M_{\odot},
\label{mb}
\ee
where $R_7$ is in unit of $10^7$ cm.
As stated above, the baryon contamination is of order 
$10^{-8} M_{\odot}~\text{to}~10^{-5} M_{\odot}$, which corresponds to 
$L_e\sim 4.47\times 10^{-5} R^{-3}_7 T_{MeV}^{-3}$ to  
$L_e\sim 4.47\times 10^{-2} R^{-3}_7 T_{MeV}^{-3}$
respectively. The thermalized fireball has a temperature 
$T\sim (L/4\pi\sigma R^2)^{1/4}\sim$ 3-10 MeV, with $L$ the luminosity and 
$\sigma$ the Stephen-Boltzmann constant. 

The processes (active to active oscillations) 
do not depend on the baryon asymmetry in the fireball, simply 
because, the neutral current contribution to the potential of all the active
neutrinos is the same and for active to active oscillation this contribution
cancels out, leaving only the dependence on $L_e$. But the active-sterile 
oscillation depends on both the lepton and the baryon
asymmetry. In Eq.(\ref{numericalpot}),  $V > 0$ for neutrino process if 
\be
L_e > 6.14\times 10^{-9} T^2_{MeV},
\label{lepasy}
\ee 
and depending on the fireball parameters and the neutrino properties
the resonance condition can be satisfied. On the 
other hand, antineutrino process will never satisfy the resonance condition
because the potential is always negative. The resonance condition of
Eq.(\ref{reso}) can be written as 
\be
L_e T^3_{MeV} =0.124 \frac{{\tilde {\Delta m^2}} \cos 2\theta}{E_{MeV}},
\label{reso2}
\ee
and the resonance length is given by
\be
L_{res}\simeq 248 \frac{E_{MeV}}{{\tilde {\Delta m^2}} \sin 2\theta}~cm.
\ee
Putting the value of $L_e$ from
Eq.(\ref{reso2}) in Eq.(\ref{lepasy}), the constraint on the 
fireball temperature is 
\be
T^5_{MeV} < 0.2\times 10^8  \frac{{\tilde {\Delta m^2}} \cos
  2\theta}{E_{MeV}}.
\label{cond}
\ee
Thus the fireball temperature derived in Eq.(\ref{cond}) is the one required 
in order to have resonant conversion of the neutrinos.
As we have already discussed, during the stellar collapse 
or merger events that trigger the burst, neutrinos of energy about 5-20 MeV 
are copiously produced
and some of these neutrinos will propagate through the fireball. 
Apart from this, due to inverse beta decay, MeV neutrinos can also be generated
within the fireball. So here we 
will take $E_{MeV}=5$ and $20$ for our calculation to estimate the fireball 
parameters.

Let us study the resonance condition
for Solar, Atmospheric and the reactor (LSND) neutrinos, where we know 
approximately the neutrino mass square differences and the neutrino mixing 
angles from the recent experimental results. These can constraint the 
fireball parameters. 
 
The recent analysis of the salt phase data of SNO\cite{snonc} 
combines with the KamLAND\cite{kam} reactor antineutrino results gives 
$6\times 10^{-5}~ eV^2 < \Delta m^2 < 10^{-4}~ eV^2$ and 
$0.8< \sin 2\theta < 0.98$ with a confidence level of 99\%. The 
best fit point has 
$\Delta m^2 \sim 7.1\times 10^{-5} ~~eV^2$ and $\sin 2 \theta\sim 0.83$.
Using the best fit point in Eq.(\ref{reso2}) we obtain
$L_e\simeq 0.5\times 10^{-5} T^{-3}_{MeV} E^{-1}_{MeV}$. 
The condition in Eq.(\ref{cond}) gives $T_{MeV} < 2.8$ and
$< 2.1$ respectively for $ E_{MeV}=5$ and $20$. 
Similarly the resonance length $L_{res}\sim 211$ kilometers and
$845$ kilometers are obtained respectively for $ E_{MeV}=5$ and $20$. 
Using the resonance value
of $L_e$ in Eq.(\ref{mb}) we obtain 
$M_b\sim 1.15\times 10^{-9} R^3_7M_{\odot} E^{-1}_{MeV}$ which is independent 
of the fireball temperature. For $ E_{MeV}=5$ and $20$ we obtain
$M_b\sim 0.23\times 10^{-9} R^3_7 M_{\odot}$ and 
$M_b\sim 0.58\times 10^{-10} R^3_7 M_{\odot}$ respectively.  
One can see that, in this case
the temperature of the fireball is less compared to the lower limit of
$3 MeV$. Also the baryon load in the fireball is less, which can be improved
by increasing $ R_7$. By considering the solar neutrino mixing angle and
the $\Delta m^2$, for neutrino oscillation in the fireball, the fireball 
temperature has to be less than $3$ MeV. If the $L_e$ vary with distance
then at the resonance $\kappa_r\sim 0.8\times 10^{-14} l_{cm} (L'_e)^{-1}$,
which implies, $L'_e=dL_e/dx$ has to be very small to satisfy the condition in 
Eq.(\ref{adbi}).

The Super-Kamiokande Collaboration recently reported the atmospheric 
neutrino oscillation parameters\cite{sk}  
$1.9\times 10^{-3} ~eV^2 <\Delta m^2 < 3.0\times 10^{-3} ~eV^2$ and 
$0.9 \le \sin^2 2 \theta \le 1.0$ with a 90\% confidence level. 
Taking $\Delta m^2\sim 2.5\times 10^{-3} ~~eV^2$ 
and $\sin^2 2 \theta\sim 0.9$ we obtain,
$L_e\simeq 0.98\times 10^{-4} T^{-3}_{MeV} E^{-1}_{MeV}$ and this implies
$T_{MeV} < 5$ and $< 3.8$ respectively for $ E_{MeV}=5$ and $20$.
So these temperature come within the range of the fireball as discussed
earlier. The resonance length for $ E_{MeV}=5$ and $20$ are respectively
$5.2$ kilometers and $21$ kilometers. Also for the above ranges of neutrino
energy, the baryon load in the fireball is 
$M_b\sim 0.44\times 10^{-8} R^3_7 M_{\odot}$ and 
$M_b\sim 0.11\times 10^{-8} R^3_7 M_{\odot}$ respectively.
Thus before coming out of the fireball, the neutrinos can have 
many resonant oscillations. If the lepton asymmetry vary with distance, then
$l_{cm}/L'_e \ge 10^{12}$ to have resonant conversion.

Thirdly let us discuss the implication of the LSND and KARMEN  
results\cite{lsnd} on the neutrino oscillation in the GRB fireball. The 
combined analysis of LSND and KARMEN 2 results give
$0.45~ eV^2 < \Delta m^2 < 1 ~eV^2 $ and 
$2\times 10^{-3} < \sin^2 2\theta < 7\times 10^{-3}$
with a confidence level of 90\%. 
We consider  $\Delta m^2 \sim  0.5~~eV^2$ and  $\sin 2 \theta\sim 0.07$ to 
estimate the fireball parameters. This gives, $T_{MeV} < 18$ and $< 14$ 
respectively for $ E_{MeV}=5$ and $20$ and for these two values of $ E_{MeV}$
the baryon load is respectively $M_b\sim 0.3\times 10^{-5} R^3_7 M_{\odot}$ 
and $M_b\sim 0.7\times 10^{-6} R^3_7 M_{\odot}$. The resonance
lengths for these two neutrino energies are respectively $0.4$ kilometer and
$1.4$ kilometer. So this shows that, the propagating neutrinos have to
oscillate several times, before they come out of the fireball. If the 
lepton asymmetry has a profile then in this case $l_{cm}/L'_e \ge 10^{10}$ to
satisfy the resonant oscillation.

If the neutrino oscillation parameters are in the  atmospheric and/or 
LSND range, the oscillation length at resonance within the fireball
will vary from few meters to $\sim$ 21 kilometers. As the
resonance length is small compared to the size of the fireball, there will be
many resonant oscillations for $\nu_{e}\leftrightarrow \nu_{\mu,\tau}$ before
they emerge out of the fireball. So the average conversation probability 
in this case is ${\cal P}(t)\sim 0.5$, where we have considered the fact that
$L_e$ does not vary with distance.
But if the oscillation parameters are
in the solar neutrino range, probably it is difficult to have neutrino
oscillation. This is due to the fact that, the fireball temperature 
in this case is less (i.e. $< 3$ MeV).

For the oscillation $\nu_{e}\leftrightarrow \nu_{s}$, with charge
neutral fireball and $L_p=L_n$, the effective potential is given by
\be
V \simeq 4.02\times 10^{-12} T^3_{MeV} 
\left [{\frac {L_e}{2}} - 6.14\times 10^{-9} T^2_{MeV}\right ]MeV.
\label{potnues}
\ee
Resonance condition can also be satisfied for the above process for 
$V > 0$. Both the processes
$\nu_{e}\leftrightarrow \nu_{\mu,\tau}$ and $\nu_{e}\leftrightarrow \nu_{s}$
are equally probable  if we consider the atmospheric neutrino oscillation
parameter or the LSND one, with average probability $\sim 0.5$.
Finally let us consider the process
${\bar\nu}_{\mu,\tau}\leftrightarrow {\bar\nu}_{s}$, for which
\be
V \simeq 2.0\times 10^{-12} T^3_{MeV} 
{L_e}~MeV.
\label{potas}
\ee
Here the higher order contribution to the neutrino potential is absent and due
to this, there is no constraint on the baryon loading of the fireball. 
Only the charge neutrality of the fireball is sufficient enough for the 
resonant oscillation to take place.

From the collapsar or merger
model of SN with SN and/or from the collision of SN with a black-hole, 
lots of neutrinos will be produced and fractions of these neutrinos will 
propagate through the fireball. 
Here we have studied the propagation of these  neutrinos through the 
GRB fireball by assuming the later to be spherical with a radius $R$, 
charge neutral and $L_p=L_n$. Also we have assumed that 
$L_e > 6.14\times 10^{-9} T^2_{MeV}$ so that the potential difference will be
positive for neutrinos and there can be resonant oscillation. By using the
known neutrino mass square difference and mixing angle from the solar, 
atmospheric and reactor experiments  we estimate the fireball temperature,
the baryon load and the lepton asymmetry in it. Our result shows that, if
the neutrino oscillation parameters are in the solar neutrino range, probably, 
there will be very few or no oscillation will take place.  On the other hand
if the neutrino oscillation
parameters are in the atmospheric and/or LSND range as discussed above, 
there can be many resonant oscillation for 
$\nu_{e}\leftrightarrow \nu_{\mu,\tau}$, $\nu_{e}\leftrightarrow \nu_{s}$ and
${\bar\nu}_{\mu,\tau}\leftrightarrow {\bar\nu}_{s}$
before they emerge out of the fireball and about 50\% of these neutrinos will
resonantly convert. 
These MeV neutrinos signals will be similar to the one from Supernovae,
for example SN1987A. Unfortunately with the present generation neutrino 
telescopes these $\nu$s can't be  detected due to 
their cosmological distance and the fluxes are extremely negligible compared 
to the galactic supernovae.

\begin{center}
{\bf ACKNOWLEDGMENTS}
\end{center}

This work is partially supported by the CONACyT (Mexico) 
grants No. 32279-E and No. 40025-F.


\begin{thebibliography}{100}
\expandafter\ifx\csname natexlab\endcsname\relax\def\natexlab#1{#1}\fi
\expandafter\ifx\csname bibnamefont\endcsname\relax
  \def\bibnamefont#1{#1}\fi
\expandafter\ifx\csname bibfnamefont\endcsname\relax
  \def\bibfnamefont#1{#1}\fi
\expandafter\ifx\csname citenamefont\endcsname\relax
  \def\citenamefont#1{#1}\fi
\expandafter\ifx\csname url\endcsname\relax
  \def\url#1{\texttt{#1}}\fi
\expandafter\ifx\csname urlprefix\endcsname\relax\def\urlprefix{URL }\fi
\providecommand{\bibinfo}[2]{#2}
\providecommand{\eprint}[2][]{\url{#2}}

\bibitem[{\citenamefont{Piran}(1999)}]{piran1}
\bibinfo{author}{\bibfnamefont{T.}~\bibnamefont{Piran}},
  \bibinfo{journal}{Phys. Rep.} \textbf{\bibinfo{volume}{314}},
  \bibinfo{pages}{575} (\bibinfo{year}{1999}).

\bibitem[{\citenamefont{Waxman}(1999)}]{waxman1}
\bibinfo{author}{\bibfnamefont{Eli}~\bibnamefont{Waxman}},
  \bibinfo{journal}{Lect. Notes in Phys.} \textbf{\bibinfo{volume}{598}},
  \bibinfo{pages}{393} (\bibinfo{year}{2003}).

\bibitem[{\citenamefont{Piran}(2000)}]{piran2}
\bibinfo{author}{\bibfnamefont{T.}~\bibnamefont{Piran}},
  \bibinfo{journal}{Phys. Rep.} \textbf{\bibinfo{volume}{333-334}},
  \bibinfo{pages}{529} (\bibinfo{year}{2000}).

\bibitem[{\citenamefont{Meszaros}(2000)}]{mes1}
\bibinfo{author}{\bibfnamefont{P.} \bibnamefont{M\'esz\'aros}}
  \bibinfo{journal}{Nucl. Phys. B (proc. Suppl.)} 
\textbf{\bibinfo{volume}{80}},
  \bibinfo{pages}{63} (\bibinfo{year}{2000}).

\bibitem[{\citenamefont{Meszaros}(2002)}]{mes2}
\bibinfo{author}{\bibfnamefont{P.} \bibnamefont{M\'esz\'aros}}
  \bibinfo{journal}{Annu. Rev. Astron. Astrophys.} 
\textbf{\bibinfo{volume}{40}},
  \bibinfo{pages}{137} (\bibinfo{year}{2002}).



\bibitem[{\citenamefont{Guilbert}(1983)\citenamefont{Guilbert, Fabian,
  and Rees}}]{guil}
  \bibinfo{author}{\bibfnamefont{P. W.} \bibnamefont{Guilbert}},
  \bibinfo{author}{\bibfnamefont{A. C.}
  \bibnamefont{Fabian}},
  \bibnamefont{and} \bibinfo{author}{\bibfnamefont{M. J.}
  \bibnamefont{Rees}},
  \bibinfo{journal}{Mon. Not. R. Astr. Soc} \textbf{\bibinfo{volume}{205}},
  \bibinfo{pages}{593} (\bibinfo{year}{1983}).


\bibitem[{\citenamefont{Derishev}(1999)\citenamefont{Derishev, Kocharovsky,
Kocharovsky}}]{deri1}
\bibinfo{author}{\bibfnamefont{E. V.}~\bibnamefont{Derishev}},
  \bibinfo{author}{\bibfnamefont{V. V.}~\bibnamefont{Kocharovsky}},
  \bibnamefont{and} \bibinfo{author}{\bibfnamefont{VI. V}~
  \bibnamefont{Kocharovsky}},
  \bibinfo{journal}{Astron. Astrophys.} \textbf{\bibinfo{volume}{345}},
  \bibinfo{pages}{L51} (\bibinfo{year}{1999}).

\bibitem[{\citenamefont{Derishev}(1999)\citenamefont{Derishev, Kocharovsky,
Kocharovsky}}]{deri2}
\bibinfo{author}{\bibfnamefont{E. V.}~\bibnamefont{Derishev}},
  \bibinfo{author}{\bibfnamefont{V. V.}~\bibnamefont{Kocharovsky}},
  \bibnamefont{and} \bibinfo{author}{\bibfnamefont{VI. V}~
  \bibnamefont{Kocharovsky}},
  \bibinfo{journal}{Astrophys. J.} \textbf{\bibinfo{volume}{521}},
  \bibinfo{pages}{640} (\bibinfo{year}{1999}).

\bibitem[{\citenamefont{Klu\'zniak}(1998)}]{klu}
\bibinfo{author}{\bibfnamefont{W.}~\bibnamefont{Klu\'zniak}},
  \bibinfo{journal}{Astrophys. J. Lett.} \textbf{\bibinfo{volume}{508}},
  \bibinfo{pages}{L29} (\bibinfo{year}{1998}).

\bibitem[{\citenamefont{Volkas}(2000)\citenamefont{Raymond, Volkas,
  and Wong}}]{vol}
\bibinfo{author}{\bibfnamefont{Raymond R.}~\bibnamefont{Volkas}},
  \bibnamefont{and} \bibinfo{author}{\bibfnamefont{Yvonne Y. Y.}~
  \bibnamefont{Wong}},
  \bibinfo{journal}{Astropart. Phys.} \textbf{\bibinfo{volume}{13}},
  \bibinfo{pages}{21} (\bibinfo{year}{2000}).

\bibitem[{\citenamefont{Paczynski}(1986)}]{pac}
\bibinfo{author}{\bibfnamefont{B.} \bibnamefont{Paczy\'nski}},
  \bibinfo{journal}{Astrophys. J. Lett.} \textbf{\bibinfo{volume}{308}},
  \bibinfo{pages}{L43} (\bibinfo{year}{1986}).

\bibitem[{\citenamefont{Goodman}(1986)}]{good}
\bibinfo{author}{\bibfnamefont{J.} \bibnamefont{Goodman}},
  \bibinfo{journal}{Astrophys. J.} \textbf{\bibinfo{volume}{308}},
  \bibinfo{pages}{L47} (\bibinfo{year}{1986}).

\bibitem[{\citenamefont{Waxman}(1997)\citenamefont{Waxman,
  and Bahcall}}]{waxman}
\bibinfo{author}{\bibfnamefont{Eli}~\bibnamefont{Waxman}},
  \bibnamefont{and} \bibinfo{author}{\bibfnamefont{J. N.}~
  \bibnamefont{Bahcall}},
  \bibinfo{journal}{Phys. Rev. Lett.} \textbf{\bibinfo{volume}{78}},
  \bibinfo{pages}{2292} (\bibinfo{year}{1997}).

\bibitem[{\citenamefont{Weiler}(1994)\citenamefont{Weiler, Simmons, Pakvasa
  and Learned}}]{weiler}
  \bibinfo{author}{\bibfnamefont{T. J.} \bibnamefont{Weiler}},
  \bibinfo{author}{\bibfnamefont{W. A.}
  \bibnamefont{Simmons}},
\bibinfo{author}{\bibfnamefont{S.} \bibnamefont{Pakvasa}},
  \bibnamefont{and} \bibinfo{author}{\bibfnamefont{J. G.}
  \bibnamefont{Learned}},
  \bibinfo{journal}{hep-ph/9411432}. 

\bibitem[{\citenamefont{Ruffert}(1999)\citenamefont{Ruffert,
  and Janka}}]{ruf}
\bibinfo{author}{\bibfnamefont{M.}~\bibnamefont{Ruffert}},
  \bibnamefont{and} \bibinfo{author}{\bibfnamefont{H.-Th.}~
  \bibnamefont{Janka}},
  \bibinfo{journal}{Astron. Astrophys.} \textbf{\bibinfo{volume}{344}},
  \bibinfo{pages}{573} (\bibinfo{year}{1999}).

\bibitem[{\citenamefont{Hjorth}(2003)\citenamefont{Hjorth et. al.,
  }}]{nature}
\bibinfo{author}{\bibfnamefont{J.}~\bibnamefont{Hjorth, el. al.}},
  \bibinfo{journal}{Nature} \textbf{\bibinfo{volume}{423}},
  \bibinfo{pages}{847} (\bibinfo{year}{2003}).
\bibitem[{\citenamefont{Raffelt}(2001)}]{raf2}
\bibinfo{author}{
\bibfnamefont{George G.}~\bibnamefont{Raffelt}},
\bibinfo{journal}{Astrophys. J.} \textbf{\bibinfo{volume}{561}},
  \bibinfo{pages}{890} (\bibinfo{year}{2001}).
  
\bibitem[{\citenamefont{Notz\"old}(1988)\citenamefont{Notz\"old,
  and Raffelt}}]{raf1}
\bibinfo{author}{\bibfnamefont{D.}~\bibnamefont{Notz\"old}},
  \bibnamefont{and} \bibinfo{author}{\bibfnamefont{G.}~
  \bibnamefont{Raffelt}},
  \bibinfo{journal}{Nucl. Phys. B} \textbf{\bibinfo{volume}{307}},
  \bibinfo{pages}{924} (\bibinfo{year}{1988}).

\bibitem[{\citenamefont{Enqvist}(1991)\citenamefont{Enqvist, Kainulaien,
  and Maalampi}}]{enq1}
\bibinfo{author}{\bibfnamefont{K.}~\bibnamefont{Enqvist}},
\bibinfo{author}{\bibfnamefont{K.}~\bibnamefont{Kainulaien}},
\bibnamefont{and} \bibinfo{author}{\bibfnamefont{J.}~
\bibnamefont{Maalampi}},
  \bibinfo{journal}{Nucl. Phys. B} \textbf{\bibinfo{volume}{349}},
  \bibinfo{pages}{754} (\bibinfo{year}{1991}).

\bibitem[{\citenamefont{D'Olivo}(1992)\citenamefont{D'Olivo,
  Torres, and Nieves}}]{jc1}
\bibinfo{author}{\bibfnamefont{J. C.}~\bibnamefont{D'Olivo}},
  \bibinfo{author}{\bibfnamefont{Manuel}~\bibnamefont{Torres}},
  \bibnamefont{and} \bibinfo{author}{\bibfnamefont{J. F.}~
  \bibnamefont{Nieves}},
  \bibinfo{journal}{Phys. Rev. D} \textbf{\bibinfo{volume}{46}},
  \bibinfo{pages}{1172} (\bibinfo{year}{1992}).

\bibitem[{\citenamefont{Sahu}(1999)}]{sahu1}
\bibinfo{author}{\bibfnamefont{S.}~\bibnamefont{Sahu}},
  \bibinfo{journal}{Phys. Rev. D} \textbf{\bibinfo{volume}{61}},
  \bibinfo{pages}{023003} (\bibinfo{year}{1999}).


\bibitem[{\citenamefont{Athar}(1995)\citenamefont{Athar,Peltoniemi, and 
  Simrnov}}]{smir}
\bibinfo{author}{\bibfnamefont{A.}~\bibnamefont{Athar}},
  \bibinfo{author}{\bibfnamefont{J. T.}~\bibnamefont{Peltoniemi}},
  \bibnamefont{and} \bibinfo{author}{\bibfnamefont{A. Yu.}~
  \bibnamefont{Smirnov}},
  \bibinfo{journal}{Phys. Rev. D} \textbf{\bibinfo{volume}{51}},
  \bibinfo{pages}{6647} (\bibinfo{year}{1995}).


\bibitem[{\citenamefont{SNO}(2004)}]{snonc}
\bibinfo{author}{\bibfnamefont{The SNO}~\bibnamefont{Collaboration}},
  \bibinfo{journal}{Phys. Rev. Lett.} \textbf{\bibinfo{volume}{92}},
  \bibinfo{pages}{181301-1} (\bibinfo{year}{2004}).

\bibitem[{\citenamefont{KamLAND}(2004)}]{kam}
\bibinfo{author}{\bibfnamefont{The KamLAND}~\bibnamefont{Collaboration}},
  \bibinfo{journal}{hep-ex/0406035}.

\bibitem[{\citenamefont{SK}(2004)}]{sk}
\bibinfo{author}{\bibfnamefont{The Super-Kamiokande}
~\bibnamefont{Collaboration}},
  \bibinfo{journal}{Phys. Rev. Lett.} \textbf{\bibinfo{volume}{93}},
  \bibinfo{pages}{10180-1} (\bibinfo{year}{2004}).

\bibitem[{\citenamefont{Church}(2002)\citenamefont{Church, Eitel, Mills
  and Steidl}}]{lsnd}
  \bibinfo{author}{\bibfnamefont{E. D.} \bibnamefont{Church}},
  \bibinfo{author}{\bibfnamefont{K.} \bibnamefont{Eitel}},
 \bibinfo{author}{\bibfnamefont{G. B.}
  \bibnamefont{Mills}},
  \bibnamefont{and} \bibinfo{author}{\bibfnamefont{M.}
  \bibnamefont{Steidl}},
  \bibinfo{journal}{Phys. Rev.} \textbf{\bibinfo{volume}{D66}},
  \bibinfo{pages}{013001} (\bibinfo{year}{2002}).


\end{thebibliography}
\end{document}